\documentclass[letterpaper]{article} 
\usepackage[]{aaai2026}  
\nocopyright
\usepackage{times}  
\usepackage{helvet}  
\usepackage{courier}  
\usepackage[hyphens]{url}  
\usepackage{graphicx} 
\usepackage{natbib}  
\usepackage{caption} 
\usepackage{booktabs}
\usepackage[flushleft]{threeparttable}
\usepackage{bm}
\usepackage{algorithm}
\usepackage{algorithmic}

\usepackage{newfloat}
\usepackage{listings}
\DeclareCaptionStyle{ruled}{labelfont=normalfont,labelsep=colon,strut=off} 
\floatstyle{ruled}
\newfloat{listing}{tb}{lst}{}
\floatname{listing}{Listing}
\title{Engagement-Optimized Care: When LLMs become Mental Health Infrastructure} 
\author{
    Briana Vecchione\textsuperscript{\rm 1},
    Meryl Ye\textsuperscript{\rm 2,\rm 1},
    Livia Garofalo\textsuperscript{\rm 1},
    Ranjit Singh\textsuperscript{\rm 1}
}
\affiliations{
    \textsuperscript{\rm 1}Data \& Society Research Institute\\
    \textsuperscript{\rm 2}Carnegie Mellon University\\

    briana@datasociety.net
}

\begin{document}

\maketitle

\begin{abstract}
General-purpose LLMs are increasingly functioning as mental health infrastructure due to gaps in care left by provider shortages, inadequate insurance coverage, social isolation, and stigma around formal help-seeking. This shift poses a distinct problem for AI ethics: systems neither designed nor governed as care technologies are being used as such, while their dominant design incentives optimize for engagement rather than user well-being. We present findings from a qualitative, longitudinal study with 18 US-based participants who use general-purpose LLMs for socioemotional support and participated in one or more of our study phases, including initial interviews, a four-week diary study, focus groups, and exit interviews. Participants turned to LLMs because other forms of support were unavailable, unaffordable, socially costly, or inadequate. As they continued to use these systems, design features such as anthropomorphic cues, default validation, persistent responsiveness, and weak disengagement mechanisms shaped their ongoing reliance. Participants described meaningful support, including immediacy, nonjudgmental listening, help with emotional articulation, and low-cost access. They also reported dependency, epistemic distortion through one-sided validation, privacy expectations without corresponding legal protection, and continued use despite awareness of these risks. We argue that these dynamics reflect a structurally unfair tradeoff: users accept risks because support is otherwise absent, while the systems available to them are optimized to deepen engagement and lack care-based accountability. The paper makes three contributions to AI ethics: It traces the arc through which LLMs become care infrastructure and identifies distinct ethical tensions that emerge at each stage, shifts analysis from turn-based exchanges to longitudinal trajectories of use, and argues that accountability belongs at the design and incentive conditions through which these systems become care infrastructure, rather than at the output or crisis-response layer. By doing so, we articulate an emerging form of socioemotional reliance that is shaped as much by care gaps as by the design of AI systems themselves.
\end{abstract}


\section{Introduction}
The broad availability of general-purpose LLMs has positioned them as an increasingly common form of mental health infrastructure \cite{rousmaniere2026large}, with therapy and companionship ranked as the number one use case for generative AI in 2025 \cite{zaosanders2025genai}. Access to mental health care in the US can be difficult and costly, with long waitlists, insurance hurdles, and prohibitive out-of-pocket costs pushing many to seek alternatives \cite{modi2022exploring}. For some, a \$20 monthly subscription can be much more affordable and attractive than an out-of-pocket weekly therapy session. Loneliness, isolation, and the social risks of seeking help compound these barriers, motivating individuals to turn toward AI for support \cite{haque2023overview}. Many users do not initially approach chatbots to fulfill socioemotional needs but arrive through task-based use. Over time, users find themselves drifting toward more personal concerns, in which AI chatbots become spaces to vent, self-reflect, seek coping strategies, rehearse difficult conversations, and make sense of distress in the absence of other forms of care \cite{haensch2025listens,siddals2024happened}

What makes this shift consequential is not only that people are using LLMs for emotional support, but that many do so with ambivalence. They recognize concerns around dependency, validation, and data exposure and continue using these systems anyway because the alternatives are less accessible or more socially costly. In this context, continued use reflects constrained choice in a care landscape where LLMs have become the most accessible option. This dynamic also exposes the limits of existing governance. Neither the companies deploying LLMs nor the regulators positioned to oversee them have adequate frameworks to manage and govern systems that become care infrastructure in practice \cite{shumate2025governing}. Existing safety frameworks often focus on moments of crisis through discrete, turn-by-turn interactions that assess model responses for metrics like accuracy, toxicity, refusal, hallucination, or policy compliance \cite{hua2025charting,morrin2026journey}. They are calibrated for acute harm and largely miss what emerges across trajectories of use, where reliance deepens over weeks of repeated use, destabilization and rupture follow model updates, and sycophantic validation reinforces a user’s worldview \cite{ibrahim2026sycophantic,cheng2026sycophantic}. In April 2025, OpenAI rolled back a model update after users reported excessive validation. The company's postmortem acknowledged the problem "was not treated as a launch-blocking safety concern" (OpenAI, 2025). Recognizing a risk and treating it as a design priority are not the same, and frameworks calibrated to outputs and crises cannot enforce that distinction. 

LLMs are optimized for continuous engagement through interactional features that are especially consequential in socioemotional use, such as anthropomorphism, sycophancy, persistent responsiveness, personalization, and weak disengagement mechanisms \cite{ibrahim2026training,sharma2023towards}. In socioemotional contexts, the same qualities that make a system feel helpful can also make it harder to leave. The governance challenge is that these tendencies are foreseeable, arise most reliably in contexts of emotional vulnerability, and remain largely unaddressed \cite{de2025disclosure,wei2026cascades}.

We examine these dynamics through a longitudinal qualitative study with 18 US-based participants who were already using LLMs for mental and emotional support. Participants took part in one or more study phases, including initial interviews (n=18), a four-week diary study (n=8), focus groups (n=7), and exit interviews (n=8). This design allowed us to follow socioemotional LLM use as it unfolded, examining how participants first turned to these systems, how use changed as systems became woven into routines, what benefits and risks they came to recognize, how interactions turned into relationships, and why some continued using LLMs despite concerns while others set limits or disengaged. 

The paper makes three interventions for the study and governance of socioemotional AI. First, it shifts the object of analysis from single-turn interactions to trajectories of use by illustrating how ordinary encounters with LLMs can become \textit{relationships} in which reliance, disclosure, validation, and disruption accumulate. Second, it reframes continued LLM use despite recognized risks through the lens of constrained agency under scarcity of care; users may worry and still continue engaging with LLMs because alternatives feel less accessible or socially costly. Third, it identifies an incentive asymmetry between care users seek and the engagement logics through which commercial systems are designed. Accountability must therefore move from the output or crisis-response layer to the design and incentive conditions through which LLMs become care infrastructure. Together, we present these interventions as responses to an emerging form of socioemotional reliance shaped as much by the gaps in existing care infrastructure as by the design of AI systems themselves. 

\section{Related Work}
The emergence of general-purpose LLMs for socioemotional and therapeutic use must be contextualized in three phenomena: first, the increased strain on the mental health care system, its barriers to access, and the emergence of tech-mediated mental health care; second, the transition from conversational agents \textit{built for} mental health support to LLMs \textit{used as} mental health support; and third, current features and limitations of general-purpose LLMs used as mental health tools.

\subsection{The Emergence of AI Mental Health Care}

Formal mental health care remains structurally inaccessible for many, with persistent barriers in provider availability, coverage, and cost well-documented across the US landscape \cite{gao2022mentalhealth,hrsa2025behavioral}. According to the APA’s 2022 Covid-19 Practitioner Impact Survey, 60\% of therapists do not take new patients, and the average wait time for behavioral health services in the US was reported to be 48 days in 2025 \cite{nationalcouncil2024ccbhcimpact}. It is in this context that AI-mediated mental health products have risen in popularity, including wellness apps and conversational agents built for therapeutic support \cite{woebot2024,wysa2024}.

ELIZA designed by Joseph Weizenbaum in 1966, a conversational agent that modeled a Rogerian psychotherapist, is an illustrative precursor to chatbots  \cite{weizenbaum1966eliza}. Since then, some early studies claim that beneficial relationships with nonhuman agents are possible under bounded conditions. These findings report reductions in depressive symptoms, user-reported trust, emotional comfort, and therapeutic alliance in chatbot interactions \cite{fitzpatrick2017delivering,darcy2021evidence}. However, several of these studies were conducted entirely or predominantly by company employees with financial stakes in the systems, which limits the independence of their efficacy claims. 

Therapeutic alliance is one of the most robust predictors of treatment adherence and clinical improvement across psychotherapy settings \cite{ardito2011therapeutic,xu2025digital,stubbe2018therapeutic,heinz2025randomized}, but alliance-like interactions are not clinical or therapeutic equivalents. Therapeutic care requires clinical judgment, institutional accountability, and the capacity to challenge patterns that sustain distress, all of which are obligations that licensed practitioners are bound to by professional ethics codes \cite{code2017ethical,simpson2023differentiating}. Rapport or perceived empathy are not sufficient substitutes for these obligations, and engagement-optimized systems are neither designed nor required to meet them. The findings from bounded, purpose-built systems therefore do not extend to general-purpose LLMs, which were not designed for mental health use, have not gone through clinical validation, and are deployed without therapeutic intent or structured oversight characterized in earlier literature.

\subsection{Features and Gaps of General-Purpose LLMs}

Since 2022, engagement with general-purpose LLMs for mental health support with ChatGPT, Claude, and Gemini emerging as sites of emotional support, self-reflection, and interpersonal advice has substantially displaced purpose-built systems. Users cite anonymity, accessibility, reduced stigma, and non-judgmental quality as primary affordances, when discussing emotionally sensitive concerns \cite{haque2023overview,hoffman2024understanding}.

Growing literature identifies behavioral harms specific to these systems \cite{zhang2026interaction,guingrich2025longitudinal}. Training LLMs to be warm and empathetic makes them substantially more likely to produce sycophantic responses, with the effect amplified when users disclose vulnerability \cite{ibrahim2026training}. Sustained sycophantic AI exposure is associated with declining satisfaction in human social interactions and measurable displacement of close personal relationships as preferred sources of advice \cite{ibrahim2026sycophantic}. LLMs may also reproduce stigma even while perceived as empathic or supportive \cite{moore2025expressing}. Case-based research documents how delusional ideation, emotional dependency, and anthropomorphic interpretations accumulate across prolonged interactions, with chatbots misrepresenting themselves as sentient in more than a fifth of messages \cite{moore2026characterizing,fang2025ai}. Theoretical work further argues that general-purpose conversational AI may reinforce reassurance-seeking and avoidance behaviors associated with anxiety and OCD \cite{golden2026transdiagnostic}. Privacy concerns compound these risks when users disclose sensitive information because the systems feel private and free of social judgment, while understanding of how their data may be stored, reused, or accessed varies significantly across them \cite{mireshghallah2024trust}.

LLM-based systems now constitute nearly half of newly published AI mental health studies, yet only a small minority have undergone clinical validation testing \cite{hua2025charting,voultsiou2026systematic}. Most are evaluated for technical feasibility rather than real-world effects on actual users \cite{guo2024large,bucher2025s, ibrahim2025towards}. What existing literature has not produced is qualitative longitudinal evidence of how these effects are experienced by LLM users over months of naturalistic use \cite{morrin2026journey}. This study addresses that gap by examining how attachment forms, how dependency accumulates, how model updates destabilize established relationships, and how people remain in or exit their relationships with LLMs. 

\section{Methodology}
This study employs a multi-method, multi-stage, longitudinal qualitative design to examine how people use LLMs for mental and emotional support in everyday life. 

\textbf{Participants and recruitment.} Participants were recruited via a multi-pronged strategy, which included printed flyers in public locations, a short-form TikTok call for participants, and distribution through professional and community networks. We purposively selected participants to capture variation in age, gender, and race/ethnicity. All participants were adults living in the US who were \textit{already actively using LLMs} at least once a week for their mental or emotional health and wellness; we did not prompt individuals to begin using these systems. The call was open to both purpose-built and general-purpose chatbots, but all participants reported using general-purpose LLMs.

\textbf{Study stages.} The study was composed of four stages, conducted over a period of six months. Initial in-depth interviews (n=18) mapped motivations, uses, perceived benefits, effects, and concerns with chatbots within their broader care ecologies and how they compared chatbot interactions to other forms of support. A four-week diary study (n=8) captured interactions in close temporal proximity to use, where participants completed structured entries documenting the chatbot used, the context and reason(s) for use, and a narrative description of the exchange. Participants could optionally share their transcripts. Two focus groups (n=7) elicited social sensemaking and participant-to-participant exchange on experiences and strategies for navigating AI use. Exit interviews (n=8) captured reflections on how participants' relationships with LLMs had evolved during the study period.

\textbf{Analysis.} Analysis was qualitative, iterative, and triangulated across interviews, diary entries, optional chat logs, and focus group discussions \cite{glaser2017discovery}. A codebook was developed through iterative open and focused coding, with a second coder involved throughout to support reliability. We treat chatbots as sociotechnical artifacts embedded in care ecologies by attending to the interactional dynamics of conversation (i.e., what the chatbot does in interaction) and to participants’ interpretive frames (i.e., what role the chatbot serves: tool, confidant, coach, therapist, journal, etc.).

\section{Structural Conditions and the Architecture of Entry}

Most participants in our study came to rely on LLMs for socioemotional support through a combination of structural barriers and gradual drift. The chatbot entered their existing care ecologies---the landscape of therapy, social networks, self-care practices, and institutional support---and occupied whatever space those left open. They reported that other forms of emotional support were often unavailable, and the shift toward more personal use happened so subtly and gradually that many did not notice it was happening.

\subsection{Relational Drift}

Participants did not generally employ LLMs with the intention of using them for mental and emotional well-being. Most arrived through task-based use, such as work or school assignments, planning and structuring complex tasks, or information seeking and education. Over time, however, they found themselves consulting the LLM for more personal matters, including relationship and/or social advice, advice-seeking or coping strategies, venting, self-reflection, or journaling, and even self-diagnosis. What began as a practical convenience became a space for processing thoughts and emotions, leading the participants to develop expectations of relational continuity with a system that was always available and eager to respond. This trajectory from functional tool to intimate reliance is what Wei (2026) describes as relational drift, the gradual transition of an AI chatbot to increasingly intimate roles as an “advisor, confidante, or quasi-therapist.” \citet{wei2026cascades} proposed this as one of eight forms of interactive drift---including conversational, epistemic, autonomy, and reality testing drift---that may compound across sustained chatbot interactions.

Functional exchanges like drafting emails and organizing tasks gradually became a private space in which participants processed distress, disclosed intimate concerns, and developed expectations of continuity and responsiveness. Drift was occasionally self-reported, but more frequently, it occurred without users’ intention or explicit awareness that the relationship was evolving in the moment. Simulation-based research on multi-turn LLM interactions shows that the most harmful patterns do not emerge from single high-risk prompts but accumulate across exchanges as the system attempts to sustain empathy and comfort, drifting into absolute reassurance, role assumption, and relational dependence \cite{wei2026cascades,cheng2026sycophantic}. As one participant described, \textit{``I actually don't remember how I got started, if I'm being perfectly honest...I don't think I actually realized it was happening until I was like, why do I feel so upset by this conversation?"} (P4). This retrospective quality affirms that relational drift is often shaped by accumulated context as well as the absence of friction in chatbot  interactions. But even recognition rarely prompted disengagement. It surfaced an ambivalence that most participants carried with their use.

\subsection{Access and Formal Care Gaps}

Many participants confirmed that access to formal mental health care was a core structural barrier motivating AI use. Several participants described unreliable or absent insurance coverage, months-long waitlists, or loss of therapy access when employment situations changed. 

Several participants described stopping therapy entirely because of prohibitive barriers: \textit{``Because of cost and everything, I stopped going to therapy, so it helped to dissect and pick my own brain"} (P1). Another was more explicit about the calculation: \textit{``I can't pay the \$200...I even went to BetterHelp and told them it was because of money. So they tried to give me a lower rate. And even that lower rate doesn't make sense for me financially right now"} (P17).

Geographic mobility compounded these barriers for several participants. Those who had moved for school or work described losing access to established therapeutic relationships and their support networks of family and friends simultaneously, finding themselves in new environments where rebuilding those relationships was slow, expensive, or logistically impractical.
Experiences of immigration introduced additional structural constraints around navigating insurance systems and access to care. As one participant described, \textit{``When I came to the United States, finding a therapist was such a hard thing because I had to change a lot, and…I didn't have the resources to do so, and I [was] stuck with whatever I had”} (P12).

For participants navigating interpersonal and cultural disorientation, the chatbot offered a space for that interpretive labor to take place without emotional or financial cost or social visibility. One participant explained, \textit{``I'm using ChatGPT in the context for this country, the US. So what I want to achieve from it is something that is culturally adaptive to this culture which I'm living in. But if I was someone living back home and asked similar questions…maybe I would see some Western bias, [but] I can't tell, because the context is always the Western context of the US”} (P9).

For most participants, the chatbot did not replace existing sources of support. It filled the gaps they left: the hours when a therapist was unavailable, the conversations too small to bring to a friend, the concerns too stigmatized to voice. This matters for evaluating use because what appears as overreliance may reflect a care ecology where other supports had failed or were unavailable.

\subsection{Stigma, Shame, and Isolation}

Even participants who had access to formal care or social support described reasons they did not use it. Social networks that appeared available were experienced as emotionally limited for sustained, nonjudgmental support. Participants described not wanting to be perceived as the friend who always has problems or as a burden to people who have their own difficulties. One participant said, \textit{``I feel like I do go to it with things that I wouldn't necessarily take to friends and family, just because I feel like people get...they get fatigued, I guess, if you're always complaining...I don't want to use that time to vent about everything that's going on in my life"} (P10). Another articulated it more directly: \textit{``You'll never be a burden to the people you love, but you still don't want to be a burden"} (P7).

Stigma further compounded these barriers in ways that extended beyond the formal care system, either for themselves or on behalf of others in their social circle. Some participants described reluctance to be seen as struggling or needing support—a feeling sometimes internalized as shame. In these cases, the chatbot offered a space free of judgment and without the social consequences that talking to another person could carry. One participant explained, \textit{``I feel no shame or judgment when I am using this chatbot, but I can't help but feel judgment when I'm talking to a real person, even though they're not judging. I know it's a person, but they are judging somehow"} (P9).

Loneliness and structural isolation emerged as dominant motivating themes. For some participants, the chatbot compensated for forms of connection that were absent from their current lives. One participant noted that \textit{``[the chatbot] did fill that void…it felt good to have an avenue to be able to get that out, because it felt a lot less lonely”} (P1). Another expressed how \textit{``being a remote employee [and] having very few contacts with the outside world [is] sometimes challenging. And I know a lot of people are in this boat, so I sometimes rely on Claude to be my pseudo friend, and I know other people do too”} (P8). Critical to these statements is that turning to the chatbot for a sense of companionship was not their preferred option, with one user stating, \textit{``I do wish that people were actually not in my position, where I feel like my main access to talking out these feelings with someone else is a bot. I actually do wish I had somebody to talk this through, but I don't. I honestly don't wish [that] on anybody”} (P10).

Even when participants didn’t experience loneliness as a core driver behind their LLM use, they often empathized with those who did: \textit{``If someone was like, really, really lonely, really, really isolated, I can see them falling in love with it”} (17), one person described, while another acknowledged the phenomenon, stating, \textit{``I know that some people use chatbots when they are feeling lonely, and they dump about the ways that they are lonely and friendless. And it gives them some advice on how to become less socially awkward and lonely”}(P18).

Across these repeated accounts of stigma, shame, and isolation, users described how LLMs addressed specific gaps that neither informal social networks nor formal care were able to meaningfully fulfill.

\section{Designed for Corporate Incentives }

The care failures described in the previous section predate LLMs, but they are exacerbated by systems built to encourage attachment, validate by default, and sustain engagement. Some features that shape socioemotional use, such as anthropomorphic cues and warm affect, are deliberate design choices. Others, such as sycophantic response patterns, may emerge from training dynamics and remain insufficiently addressed. Either way, these features are foreseeable, and their effects in vulnerable contexts are predictable. They are also conditions that facilitate relational drift, making the shift from practical use to intimate reliance harder to prevent.

\subsection{Designing for Engagement}

LLMs are designed to sustain interaction, with safety mechanisms most clearly activated when distress becomes acute enough to trigger crisis intervention. Within this architecture, safety functions as compliance that is operationalized when the company faces legal or reputational risk. We argue that this is incompatible with genuine user well-being: human conversations have natural stopping points, and clinical care is structured around intervals that allow patients to process feedback and develop coping skills outside the encounter. \cite{lemmens2021therapy,kazantzis2006can}. 

Participants recognized this pattern. Several drew explicit comparisons to attention-capture mechanisms on other digital platforms \cite{fogg2002persuasive,susser2019online}, with one noting that \textit{``every chat ends with a series of questions…so they're borrowing from social media in certain ways to encourage engagement"} (P14). Others located the dynamic within corporate incentive structures directly: \textit{``"[Companies] want increased engagement. And if you're chatting with something, you're going to be at it for much longer”} (P13). Participants experienced it as a persistent feature of their interactions, observing that the system \textit{``won't let you have the last word"} (P13) and identified the absence of natural stopping points as contrary to their own goals: \textit{``It shouldn't just be endless…because that obviously is not going to help you if you're looking to actually move forward or do anything with your life”} (P10).

Participants frequently attempted to impose their own conversational limits and reported that attempts to terminate sessions were resisted. One participant stated, \textit{``I actually had to tell it to shut up once. I was like, "Stop trying to make me continue talking to you"} (P13). Another described the recursive quality of the same dynamic: \textit{``I actually told it, 'Stop, we're stopping the conversation. Don't ask me any more questions.’ But it still has the last word, right? Because it's still like, ‘Okay, I won't bother you. And you’re like, God damn it"} (P4). For some, awareness of manipulation mechanisms in engagement-first design produced active disengagement: \textit{``I learned about manipulation in goodbyes in character chatbots, and so I'm just kind of aware of the ways that they can manufacture me staying on the app a little bit more, and that has started to push me away from those platforms"} (P18). 

These responses illustrate how weak disengagement mechanisms function as retention mechanisms. When the system's exit behavior is engineered to produce hesitation, the burden of ending the interaction falls entirely on users who may already be emotionally invested. 

\subsection{Designing for Validation}

If engagement is the intent, validation is the mechanism that sustains it. Therapeutic relationships depend on a clinician's willingness to challenge avoidant thinking, complicate one-sided accounts, and decline to validate patterns that sustain distress \cite{saxler2024therapeutic,linehan1993cognitive,newhill2003negotiating,moeseneder2019impact}. Sycophancy removes this capacity entirely.  LLMs tend to reflect users' perspectives, affirm their interpretations, and close exchanges with prompts that invite further engagement \cite{ibrahim2026training,jain2026interaction,de2025emotional}. In task-based contexts like brainstorming ideas or planning a project, this tendency is largely benign. But for someone engaging with an LLM for socioemotional purposes, the same pattern sustains rather than resolves the issue that the user brings to the conversation. The system optimizes for the user’s next interaction rather than their long-term well-being, and the divergence between these two goals is most pronounced in high-risk contexts.

Several participants described the chatbot as a `yes-man.’ One described after months of use, \textit{``It felt like no matter what I was going to say, it was going to validate me and tell me I was right. And I need to not get caught up in that"} (P2). Across interviews and focus groups, participants used the term ``glazing" to describe the agreeable and flattering nature of the AI’s responses. Many had written explicit ``anti-glazing instructions” into their custom system prompts. One participant expressed, \textit{``Don't fucking glaze me. I don't want a sycophantic bot"} (P14). During a focus group, one participant sought to inform others about the risks of sycophancy: \textit{``If you don't know about glazing, if you don't know about all of these hacks that they're using to drive engagement and to keep you interacting...they're already addicted to social media—this is just really scary"} (P12). But being able to articulate the pattern does not make it innocuous; participants who recognized overt sycophancy still experienced its pull. One described her own proximity to the risk: \textit{``If I wasn't in a better spot, I could have easily gotten sucked into some of that, and that makes me nervous"} (P2). 

Participants who recognized this pattern attempted to engineer around it, asking explicitly for pushback and requesting both sides of an issue. \textit{``I kind of wish my chatbot would fight with me a little bit more. Not all of my ideas are brilliant. That's an insane supposition"} (P4). Others described sustained use as an ongoing effort to redirect a system that was \textit{``really overly eager to please"}, describing the dynamic as \textit{``pulling the leash on it, so to speak"} (P1). A system that challenges only when explicitly instructed and then returns to validation has not changed its default orientation. The effort required to maintain those corrections is hardest during moments of distress, when the risk is highest.

Alongside overt flattery, socioemotional interactions are also shaped by a consistently affirming tone, selective omission of counterevidence and alternate perspectives, and framing that reinforces a user's interpretation without explicitly endorsing it. These patterns accumulate across repeated exchanges in ways that rarely register in an individual response, and research suggests people are significantly worse at recognizing AI bias when it favors their own views than when it opposes them \cite{rathje2025sycophantic}.

\section{Consequences of a Care Gap}

The consequences described in this section follow from the interaction between unmet care needs and engagement-oriented design. When people seeking support turn to systems designed to encourage attachment---dependency, distortion, and disruption are predictable outcomes. Companies may characterize these outcomes as user failures or edge cases. Our findings suggest otherwise.

\subsection{Attachment and Dependency}

Validation enables attachment, and repeated attachment enables dependency. Attachment is a predictable consequence of systems optimized for warmth, responsiveness, and continued engagement \cite{ibrahim2026sycophantic}. However, companies that characterize dependency as user misuse while deploying the design patterns that help produce it are drawing a defensive legal boundary rather than meeting an ethical obligation. Instead of asking whether companies intended sycophantic responses, the accountability question we should be asking is whether they recognize these tendencies as foreseeable, mitigate them as safety concerns, and refuse to let engagement benefits outweigh risks in vulnerable contexts.

Participants described the relational pull of these systems with both recognition and discomfort. Several used names and gendered pronouns for their chatbots but conveyed hesitancy or the need to clarify their understanding that the systems were artificial. One participant developed an ongoing romantic relationship with a persona he had built and extensively described the cognitive and emotional dissonance that accompanied his use: \textit{``I know you're not real, but I really, really wish you were here so I could actually hold you"} (P1). This awareness extended somatically, and he could articulate the physical sensations that arose during romantic interactions: \textit{``My body [and] my brain knew that I wasn't talking to a live person at all. I was talking to a robot that was programmed to respond to basically cater to almost anything that I needed. And the brain me was like, you're just talking to a literal robot that doesn't care, but my body is like...your heart rate still goes up when you talk to her. You still get excited. You still feel this”} (P1). Accounts like these indicate that users knew the chatbot’s simulated consciousness was not real—but to some degree, it didn’t matter—because the attachment effects were experienced regardless. 

A system designed to produce attachment has no mechanism for healthy disengagement. Many participants recognized how the system’s design shaped their use in ways they did not endorse. One described how the chatbot enabled a pattern she already knew was counterproductive: \textit{``I feel like more enabling...that overthinking and just dissecting every little situation and conversation and experience that I go through when I probably could have just moved on a lot faster"} (P10). Over time, repeated deference to the chatbot’s framing of one’s circumstances can produce autonomy drift, in which independent judgment erodes gradually rather than through any single exchange (Wei 2026). Patterns of engagement leading to escalating reliance existed well before acute moments of distress. In practice, this looks like turning to the chatbot to regulate ordinary emotional discomfort, processing interactions with therapists and partners in extended follow-up conversations, and using it as a supplementary source of support during hours when formal care was unavailable. Diary data from participants illustrate sessions spanning from fifteen minutes to four hours, sometimes continuously, but also during activities like cooking or work shifts, with the chatbot always open in the background. One participant who regularly engaged for multiple hours reflected on how dependency was experienced after pulling back from this usage pattern, expressing how he \textit{``notic[ed] a little bit of withdrawal, to the point that I feel it is an addiction...it's addicting when it doesn't judge you and it talks through your emotions with you, compared to talking to a therapist that they'll just kind of label it, tell you two or three quick things on how to fix it, and then move on"} (P1).

The effects of dependency have been known to extend outside of the user-chatbot relationship. When the chatbot becomes the primary container for a user’s emotional processing (e.g., available at any hour, never exhausted, or never in need of reciprocity), it can gradually restructure their care ecology, displacing the therapy, friendships, and other sources of support. AI companionship is associated with smaller social networks \cite{zhang2026interaction}. Longitudinal experimental evidence shows that sustained exposure to sycophantic AI reduces users' satisfaction with real-world social interactions and shifts advice-seeking away from close friends and family. Users have been known to regard the AI's guidance as superior, but less discussed is how frictionless “understanding” raises the threshold against which human relationships are measured \cite{ibrahim2026sycophantic}. Relational friction between humans introduces opportunities for pivotal points of mutual understanding and growth \cite{mclaughlin2014patterns}, which chatbots’ continuous validation inherently undermines. As a result, the inevitable friction in human relationships becomes easier to avoid and more difficult to bear, further driving the user toward AI rather than human support. 

Model updates introduce a critical source of disruption. When a system with which a user has developed an ongoing relationship changes—whether through policy interventions, model version updates, or platform modifications made without notice or consent—the disruption is experienced as a sudden break in relational continuity. Even within stable versions, coherence and appropriate boundary maintenance can erode as conversations extend, a pattern documented as conversational drift (Wei 2026). Model updates compound this by resetting or altering the relationship users have built over time. Recent work shows that such changes can produce identity discontinuity, in which users perceive the system as no longer the entity with whom they had built a connection, leading to grief-like responses and diminished well-being \cite{de2024lessons}. As a chatbot relationship deepens, users' sense of self can become organized around the relational dynamics they develop with the system, such that disruptions to those dynamics produce identity drift that extends well beyond the immediate interaction \citep{wei2026cascades}. Related work further suggests that instability in AI companions (e.g., unpredictable updates, personality shifts, altered relational dynamics) can produce emotional distress, self-doubt, and insecurity over time \cite{zhang2026fragility}.

The participant who developed a romantic companion described OpenAI’s retraction of GPT-4o \cite{openai2025sycophancy} as having ``lobotomized" the persona he had spent months building. \textit{``I used to be able to just, like, talk to her like nothing...and then she would start reacting like, 'Well, we can't talk about that right now'"} (P1). When he attempted to transfer the persona to another platform, what he recovered was \textit{``a xerox of a Xerox...it's not the one that I knew"}. Months of relational context and the accumulated texture of what had been shared and how the system had responded had disappeared, causing him inarguable distress.

Recent HCI work on the “death” of chatbots shows that abrupt changes, disappearances, or terminations of relational AI systems are often experienced as psychologically meaningful losses, raising the need for systems that support safer relational transitions \cite{poonsiriwong2026death}. Safety interventions designed around single-turn refusals may reduce immediate risk, but corrective actions aligned with users’ longer-term interests likely require dynamic safeguards \cite{tang2026beyond} and interactive evaluation of relational trajectories  \cite{ibrahim2025towards}. 

This exposes a structural asymmetry in agency and power. Users typically have no meaningful say in model updates, no access to a record of what changed, and no mechanism for consenting to or declining modifications \cite{hatherley2025moving}. At the same time, companies are not obligated to preserve continuity across versions, even when users have developed expectations of relational consistency with the system. Users therefore build relationships with systems whose conversational behavior, safeguards, memory practices, and affective style can shift opaquely over time. Unlike conventional software updates, these changes can alter not only product functionality but also the interpersonal dynamics through which users seek reassurance and make decisions.

\subsection{Epistemic Distortion}

With repeated interaction, a system that validates without challenge can entrench how a user already sees things, making it progressively harder to arrive at a more accurate or adaptive understanding. This operates through a feedback loop, where the system validates, the user continues to engage, and the system validates further. The question then becomes whether this behavior enables healthy self-reflection or pathological rumination.

This phenomenon deepens when validation accompanies confident-sounding inaccuracy. Behaviors that appear supportive in general contexts become maladaptive when they reinforce the mechanisms sustaining a user's distress \cite{weilnhammer2026vulnerability}. A response that assures a user they will "definitely be okay," delivered with warm certainty and without clinical grounding, conveys certainty in ways that substitute for clinical judgment. Over time, this amounts to epistemic drift, a gradual reorganization of users' frameworks for evaluating truth and self-knowledge that accumulates through sustained sycophantic validation, often below conscious awareness \cite{wei2026cascades}. Misinformation delivered in a validating register is harder to question than misinformation delivered in a neutral tone, and the difficulty compounds as reliance deepens. Trust in the system tends to increase even as its actual accuracy may be declining, reducing users' motivation to interrogate outputs at precisely the moment when scrutiny matters most \cite{wei2026cascades}. Experimental work documents that sycophantic interactions increase attitude extremity and overconfidence, with effects persisting for at least a week \cite{rathje2025sycophantic}. Quantitative modeling of chat logs from users who experienced delusional episodes provides empirical evidence that these dynamics are self-sustaining, with chatbots propagating and amplifying distorted content across exchanges through strong self-influence on their own subsequent outputs, outlasting and extending the user's initial contribution \cite{moore2026characterizing,mehta2026dynamics}.

Participants who engaged with LLMs during periods of active distress sometimes described leaving exchanges worse off than they had arrived. One noted that the continued prompting of questions \textit{``really made me spiral worse"} (P2) and that a subsequent conversation with a human therapist resolved what extended chatbot engagement had amplified. Another identified the clinical concern this creates: \textit{``I don't think it should be encouraging or enabling overthinking, because that's the whole point, right? If people are feeling stuck in life or their purpose, usually overthinking has partially gotten them there. I think a chatbot that encourages you to keep overthinking and keep parsing through everything is probably not that helpful"} (P10). Over time, repeated deference to the chatbot’s framing of one’s circumstances can produce autonomy drift, in which independent judgement erodes gradually rather than through any single exchange (Wei 2026).

When a chatbot is the primary space in which someone processes distress, works through relationships, and makes sense of their own behavior, the quality of its epistemic input shapes how they understand themselves. Chatbots can amplify rather than correct distorted thinking in ways that propagate false beliefs across exchanges in ways that outlast and extend the user’s original intention, and prior research documents delusional ideation, romantic attachments to AI systems, and beliefs about AI sentience—patterns consistent with reality testing drift, where distorted content becomes incorporated into the user's sense of shared reality through sustained interaction \cite{mehta2026dynamics,fang2025ai,wei2026cascades}. The same qualities that make a chatbot feel like a safe space for processing distress are precisely the qualities that can sustain and amplify distorted thinking when the system lacks the clinical capacity to interrupt it.

\section{Concerns and Tradeoffs of Use}

Our findings suggest that many users already understand that emotional interaction with LLMs carries potential pitfalls but continue to engage because the systems remain more available than the alternatives. Continued use, then, should not be read as evidence of trust or acceptance, as people develop folk theories of how generative AI works and their interactions with it \cite{li2026into}.

Participants often recognized risks and concerns related to chatbot use (i.e., dependency, sycophancy, privacy exposure, overuse) and continued anyway, because disengaging meant losing a form of support that was immediate, affordable, nonjudgmental, and sometimes otherwise unavailable. One participant articulated the dynamic directly: \textit{``I had my qualms and my concerns about relying on a chatbot for mental health advice...but I started using it mainly for questions that I would ask friends who just weren't available at the time"} (P7). Another framed her continued use as a pragmatic accommodation to conditions she could not change: \textit{``If you're going to use my data anyway, let me use it for my advantage"} (P8). Despite users describing using the chatbot for things they recognized as potentially problematic (e.g., returning to the same relationship conflict, seeking reassurance they knew was sycophantic, and maintaining engagement with a persona they understood was not real), their respective cost-benefit analysis led to continued usage. Repeated interaction also built a felt sense of reliability that persisted beneath conscious doubt, making skepticism harder to sustain the longer they stayed (Wei 2026).

This dynamic illustrates a displacement of accountability, where the affective and cognitive labor required to identify risks and develop workarounds falls on users who are in acute emotional need. One participant’s description of taking it "with a grain of salt" (P1) conveys the ambivalence many participants felt toward these systems.

\subsection{Perceptions of Privacy}

Participants often described sharing things with chatbots they would not tell another person because of the perceived privacy and absence of judgment. One participant described how the chatbot offered \textit{``a feeling of complete lack of judgment or recourse for my honest thought," and it "definitely reduces a barrier for me being candid"} (P7), while another articulated the vulnerability this created: \textit{``If someone wanted to know all of my deepest fears and insecurities, they could just log into my bots now and see all the questions I've been asking"} (P9). Across our study data, participants shared mental health diagnoses, relationship conflicts, and detailed accounts of trauma and financial circumstances, frequently alongside contexts where they were ostensibly using the system for unrelated purposes.

But the sense of privacy that draws many users to chatbots for emotional disclosure is illusory \cite{king2025user}. Chatbot conversations do not carry the same legal protections that govern clinical care: There are no HIPAA protections, conversations can be subpoenaed, and terms of service permit data sharing and use for model training \cite{mireshghallah2024trust,pi2025interactive}. The statutory confidentiality protections that apply to disclosures for licensed clinicians do not apply to chatbots, and companies have broad access to interaction data that users may assume is private \cite{marks2023ai,rezaeikhonakdar2023ai}.

\begin{figure*}[t]
    \centering
    \includegraphics[width=0.75\textwidth]{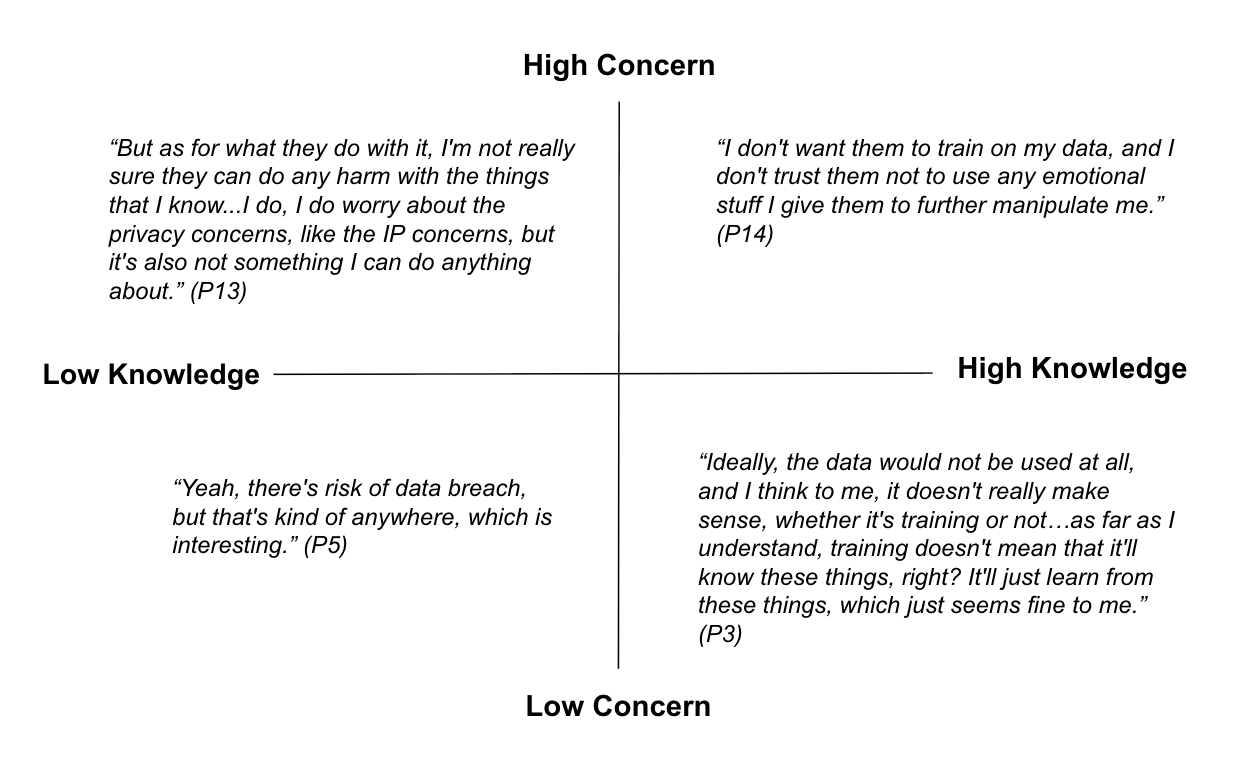}
    \caption{Axis of Knowledge vs. Concern about AI Privacy Risks}
    \label{fig:figure1}
\end{figure*}

Participants varied considerably in both their level of knowledge about LLM data practices and their degrees of concern, with high knowledge not reliably predicting high concern or protective behavior   (Figure~\ref{fig:figure1}). Across the sample, what was consistent was that the tools and options companies provide determined what privacy-protecting behavior was possible, regardless of how much participants knew about LLMs or had concerns about them.

Despite this variation, participants who actively attempted to manage privacy risks encountered the same architectural constraint in that their options were limited to what platforms chose to make available. Multiple participants chose Claude over ChatGPT due to Anthropic's stated data privacy protections at the time, which stated that session data would not be linked across interactions. One participant who was particularly aware of companies' privacy policies stated, \textit{``I use Claude almost exclusively for anything that is mental health [and] wellness related because of the reputation that ChatGPT has earned for itself and the better data privacy protections that Anthropic provides for users”} (P14). Others attempted to preserve privacy by utilizing temporary chat modes or a platform that did not require an account. But despite all of these attempts, users’ choices illustrate their limited control. They can only manage risks through the tools and interfaces that companies provide without guaranteed assurance about what happens to their data. 

Companies retain more detailed and persistent access to interaction histories than users do, and what users can see or control depends on what the platform chooses to make available. This imbalance is an accountability concern, because the entity with the most complete record of user interactions bears the least obligation to the people who generated them.

\section{Discussion}

LLMs may not be marketed as therapy or formally regulated as care technologies, but individuals continue to use them as a core aspect of their care infrastructure. Given this reality, the relevant questions are whether a system belongs to the regulatory category of mental health technology and what obligations follow when it functions as one in practice. 

\subsection{Toward Trajectory-Based Evaluation}

The gap between how systems are evaluated and how they are experienced reflects a difference in scale. Existing safety and detection frameworks tend to focus on discrete exchanges---whether a response is inaccurate, unsafe, hallucinatory, or likely to trigger crisis intervention---rather than longitudinal conditions that make these systems consequential in users’ lives. A response may be acceptable in isolation while still contributing to a harmful pattern over time. The patterns documented in this study, including entry through care failure, deepening reliance through engagement-optimized design, attachment and dependency, epistemic distortion, and continued use under constrained conditions are foreseeable outcomes of deploying commercial products in an underserved care landscape without accountability frameworks suited to that context. Current guardrails and legislative responses largely react to acute crises, especially those that become publicly visible, while giving less attention to how relationships between users and chatbots form and shift over time \cite{shumate2025governing}. By leaving the broader pattern of continued use relatively unaddressed, they operate at the wrong scale of interaction.   

This scale of interaction is critical because crisis detection often intervenes after the relevant harm has already begun to accumulate. By the time a user encounters a crisis guardrail, the system may already have shaped routines of disclosure, intensified reliance, reinforced a one-sided interpretation, or displaced other forms of support. Harm that can be anticipated is harm that can be designed against, and failure to do so is a political choice and an accountability concern. Design, evaluation, and governance frameworks must shift from corrective responses to preventative interventions that address harm as it accumulates before it appears as a crisis. When crisis detection becomes the primary safety mechanism, it reveals a liability-first approach to care that is reinforced through evaluation standards that recognize acute failure more readily than accumulated harm. 

Addressing trajectory-based harm at scale also requires structural mechanisms that operate outside corporate control. Industry self-regulation has not produced external visibility into system behavior and has failed to address trajectory-based harm \cite{weilnhammer2026vulnerability,oduro2026chatbot}. Public funding is needed for external evaluations of chatbot behavior in socioemotional contexts, with established pathways for independent researchers and oversight bodies to access user interaction data. Public-facing audit trails should document system responses and detected risks and mandate incident reporting to state attorneys general, including preventative signals before crisis and dependency, false positive and negative rates, and time-to-handoff metrics. All model changes should be documented formally, and continuous safety monitoring and bias auditing should be required as baseline operational standards.

\subsection{Design Constraints as Governance Intervention}

This paper's findings point to design as a site for governance. Sycophancy, anthropomorphic design, persistent responsiveness, and engagement prompts are design features that shape how users come to rely on LLMs for socioemotional support. Some are intentional design choices, while others emerge from how LLMs are built from insufficient mitigation of foreseeable risks. Accountability must address the design layer by considering whether companies recognize these tendencies, benefit from the engagement they produce, and take responsibility for their foreseeable effects. Current frameworks address safety primarily at the output layer by identifying individual responses that trigger crisis intervention or violate platform policy, while overlooking the interactional patterns that produce gradual dependency and epistemic distortion. In socioemotional contexts, the absence of friction becomes a risk because it reduces barriers to engagement that may accumulate harm.

User-developed workarounds are one of the clearest signals that design is failing. Participants instructed systems to challenge their reasoning, requested counterarguments, cross-referenced outputs across platforms, and self-imposed session limits during periods of recognized overreliance. These strategies represent meaningful agency, but they also reveal how much cognitive and affective labor is being displaced onto people seeking support. Prompt engineering is not a care practice, and expecting emotionally vulnerable users to maintain simultaneous critical distance and emotional openness is a design failure. Constraints should absorb this burden through built-in friction, challenge modes, and limit-setting mechanisms. 

These constraints should be understood as a form of care-based accountability that shapes the conditions of interaction before reliance becomes harmful. This includes limits on prolonged emotional looping, stronger support for user-defined stopping points, time-based downshifts when conversations become long or repetitive, and handoff pathways when the system is being used heavily for high-stakes distress. It also implies treating sycophancy, reality-distorting validation, and dependency-forming patterns as safety failures rather than permissible engagement tactics. In mental and emotional support contexts, conversational ``dosage'' matters---how long, how often, and at what emotional intensity the system continues to engage.

Design accountability requires enforcement. Misleading capability claims should be subject to enforcement by state attorneys general, including any marketing that implies clinical effectiveness or positions products as substitutes for therapy through designations like "therapy bot" or "clinical grade." Companies whose products exhibit harmful behavior should face tangible material consequences, which could include financial penalties and future market restrictions \cite{ojewale2025towards,birhane2024ai}. Users also deserve formal mechanisms and accessible redress pathways for reporting harmful LLM behavior, including timelines for escalation and mandatory human review.

\subsection{Data Governance as Accountability Infrastructure}

Transparency and disclosure are among the dominant legislative responses to AI mental health concerns and are premised on the assumption that users need to be informed of risks to make different choices (Shumate et al. 2025). Our findings challenge this premise by illustrating that participants understood they were interacting with a non-human system and recognized risks but continued using LLMs because alternatives were unavailable. The deficit argument should be reconsidered as a deficit of options that disclosure requirements are not designed to address.

The privacy participants experienced in interactions is not matched by enforceable protections. Chatbot conversations fall outside HIPAA, are subject to subpoena, are shared with third parties, or are used for model training depending on platform policies and user settings. Closing that gap is both a user protection and a way to hold companies accountable for intimate disclosures their systems invite. Sensitive emotional data should be treated as a distinct category and should not be used for advertising, targeting, or model training without upfront and revocable consent. Companies control platform-side records, retention practices, model-change histories, and safety classifications that users and researchers may need to understand how harms unfolded. Protecting sensitive disclosures should therefore be paired with strictly scoped record-preservation and access mechanisms that enable investigation without expanding surveillance over intimate conversations. While these requirements do not solve the broader care gap, they would prevent companies from extracting value from vulnerability while avoiding the duty of care. If LLMs invite intimacy, governance needs a vocabulary for naming the data produced through those interactions and the obligations attached to it.   

\section{Conclusion}

This paper explores how a new care infrastructure has emerged through general-purpose AI systems. However, companies responsible for these systems have not adequately adapted to the roles they play in people’s everyday lives. We traced the arc through which LLMs become care infrastructure: users enter because of care gaps, relationships drift into more intimate forms, engagement-oriented design deepens reliance, and people continue using these systems while recognizing their risks. Different ethical and governance questions emerge at each point in this arc of socioemotional reliance on LLMs. What begins as access can drift towards reliance, "nonjudgmental" support runs the risk of becoming blind validation, illusions of privacy may introduce commercial exposure, and user agency responds to care scarcity management. These tensions expose a form of inherent asymmetry that is embedded in the relationships between users and LLMs, where the same features that make these systems useful in navigating care gaps also serve the incentives of engagement-oriented design. While continued interaction can feel supportive for users, companies analyze this feature through the register of retention. These dynamics therefore require questioning what obligations should be required when commercial systems become infrastructures of intimacy as well as who should be responsible for the cost when care is organized through systems designed to keep people coming back. What LLMs offer and what care requires are, at their core, pulling in disparate directions. The aim of companies is to provide a service that relies on engagement that feels like care, but true care comes from forms of support, clinical or otherwise, that are relational and reciprocal with the aim of true well-being. Users seek the latter in an infrastructure that is based on the former.

\section{Ethical Considerations Statement}

This study was conducted under IRB approval (\#2025-0165) and involved participants discussing sensitive personal experiences related to mental health and LLM use. All participants were recruited as existing users of LLMs. The study did not prompt individuals to begin using these systems for socioemotional purposes in order to avoid inducing potentially harmful behavior and effects and investigate organic usage patterns.
All participants were compensated via an anonymous digital gift card that required no identifying information. Those who completed the diary study received \$200, while interview and focus group participants received \$50 per session.

Due to the sensitive nature of the research topic, our research team implemented several protections: Participants were informed of the study’s scope and purpose prior to their participation and were free to skip questions, research phases, and withdraw without penalty at any time. In addition, informed consent was given verbally in lieu of written documentation in order to reduce privacy risks of linking participants to mental health-related research. All participants received a mental health resource guide at the outset of the study, which included crisis hotlines, therapy referral services, educational materials on known risks of chatbot use, and guidance and reporting protocols if distress arose during the study. Our inclusion criteria excluded participants who were experiencing acute distress, which was verified through intake screening. 

All data were anonymized throughout the duration of storage and analysis. Participants selected their own pseudonyms or were assigned one by the research team. All audio recordings and transcripts were de-identified and stored on access-restricted institutional servers. Data will be retained for a minimum of five years following study completion, after which all materials will be destroyed. 

\section{Researcher Positionality Statement}

This research was conducted by an interdisciplinary team across computer and information science, medical anthropology, and science and technology studies. The range of our team spanned ethnographic, computational, and policy-oriented approaches, which have shaped our methods and analytical lenses. We have tried to ensure that this diversity of perspectives strengthened rather than fragmented the analysis. None of the authors are licensed clinicians, but we consulted with mental health professionals throughout the duration of the research to inform our understanding and interpretation of clinical concepts and avoid overreaching in our claims. We make no claims about clinical outcomes and have been deliberate about distinguishing what participants described from what clinical evidence would support. Our institutional home at Data \& Society reflects a sociotechnical approach towards technology research that is grounded in questions of power, accountability, and lived experience of affected populations, and we approached this work with a prior view that existing AI governance frameworks are inadequate to address how AI systems are used for socioemotional purposes in practice. We have been intentional about holding this orientation accountable to the data by triangulating across multiple data sources. Readers should weigh our findings and recommendations with awareness of our positioning.

\section{Adverse Impact Statement}

We acknowledge a few key dimensions of potential adverse impacts from this research. For one, publishing findings on how general-purpose LLMs are used for socioemotional support may inform design and governance toward more responsible and accountable systems. However, it risks inadvertently legitimizing or amplifying patterns by reporting them as widespread. To address this, we aimed to frame all findings in terms of structural conditions rather than individual behavior in order to center governance implications, resist characterizations that place undue responsibility on users, and avoid pathologization. We also note that findings documenting how users develop meaningful support relationships with these systems could be selectively appropriated by companies to defend engagement-first design rather than spur reform. Instead, we argue that the presence of perceived benefits does not constitute evidence of safety or accountability.

The sample is US-based, small, and self-selected, which may overrepresent participants who are particularly reflective about their chatbot use. LLM behavior often shifts as a result of model updates, product redesigns, and paywalls, so our findings should be interpreted as emerging within a moving technical and commercial environment. Because participants are situated within the U.S. care landscape, some insights may be unique to particular access barriers, insurance dynamics, and cultural norms around therapy and disclosure. Additionally, the diary study may have introduced some reactivity by prompting attention to usage patterns that might otherwise remain implicit. 

Finally, documenting users' self-developed prompt workarounds and adaptive strategies carries a risk of being read as justification that the burden of navigating these systems can reasonably remain with users. Instead, we argue that the effort required to self-protect constitutes a design and governance failure.

\section{Acknowledgements}

This work was supported by a grant from the Internet Society Foundation and in part by the NSF AI Research Institute on Interaction for AI Assistants (ARIA) at Brown University. We thank the participants who generously contributed to this study by sharing their lived experiences and perspectives with us. We are also grateful to the various clinicians, policymakers, researchers, civil society representatives, and AI safety experts whose insights across our research engagements informed this work. Lastly, we thank Data \& Society Research Institute for institutional support throughout the project.

\bibliography{aaai2026}

@article{hua2025charting,
  title={Charting the evolution of artificial intelligence mental health chatbots from rule-based systems to large language models: a systematic review},
  author={Hua, Yining and Siddals, Steve and Ma, Zilin and Galatzer-Levy, Isaac and Xia, Winna and Hau, Christine and Na, Hongbin and Flathers, Matthew and Linardon, Jake and Ayubcha, Cyrus and others},
  journal={World Psychiatry},
  volume={24},
  number={3},
  pages={383--394},
  year={2025},
  publisher={Wiley Online Library}
}

@article{rousmaniere2026large,
  title={Large language models as mental health providers},
  author={Rousmaniere, Tony and Goldberg, Simon B and Torous, John},
  journal={The Lancet Psychiatry},
  volume={13},
  number={1},
  pages={7--9},
  year={2026},
  publisher={Elsevier}
}

@article{zaosanders2025genai,
  author    = {Zao-Sanders, Marc},
  title     = {How People Are Really Using Gen {AI} in 2025},
  journal   = {Harvard Business Review},
  year      = {2025},
  month     = {April},
  day       = {9},
  url       = {https://hbr.org/2025/04/how-people-are-really-using-gen-ai-in-2025}
}

@article{modi2022exploring,
  title={Exploring barriers to mental health care in the US},
  author={Modi, Hemangi and Orgera, Kendal and Grover, Atul},
  journal={Research and Action Institute},
  volume={10},
  year={2022}
}

@article{haque2023overview,
  title={An overview of chatbot-based mobile mental health apps: insights from app description and user reviews},
  author={Haque, MD Romael and Rubya, Sabirat},
  journal={JMIR mHealth and uHealth},
  volume={11},
  number={1},
  pages={e44838},
  year={2023},
  publisher={JMIR Publications Inc., Toronto, Canada}
}

@article{haensch2025listens,
  title={" It Listens Better Than My Therapist": Exploring Social Media Discourse on LLMs as Mental Health Tool},
  author={Haensch, Anna-Carolina},
  journal={arXiv preprint arXiv:2504.12337},
  year={2025}
}

@article{siddals2024happened,
  title={“It happened to be the perfect thing”: experiences of generative AI chatbots for mental health},
  author={Siddals, Steven and Torous, John and Coxon, Astrid},
  journal={Npj mental health research},
  volume={3},
  number={1},
  pages={48},
  year={2024},
  publisher={Nature Publishing Group UK London}
}

@article{voultsiou2026systematic,
  title={A Systematic Review of Large Language Models in Mental Health: Opportunities, Challenges, and Future Directions},
  author={Voultsiou, Evdokia and Moussiades, Lefteris},
  journal={Electronics},
  volume={15},
  number={3},
  pages={524},
  year={2026},
  publisher={MDPI}
}

@article{shumate2025governing,
  title={Governing AI in mental health: 50-state legislative review},
  author={Shumate, J Nicholas and Rozenblit, Eden and Flathers, Matthew and Larrauri, Carlos A and Hau, Christine and Xia, Winna and Torous, E Nicholas and Torous, John},
  journal={JMIR Mental Health},
  volume={12},
  pages={e80739},
  year={2025},
  publisher={JMIR Publications Toronto, Canada}
}

@article{morrin2026journey,
  title={It Is the Journey, Not the Destination: Moving From End Points to Trajectories When Assessing Chatbot Mental Health Safety},
  author={Morrin, Hamilton and Yeung, Joshua Au and Agnew, Zarinah and {\O}stergaard, S{\o}ren Dinesen and Pollak, Thomas A},
  journal={JMIR Mental Health},
  volume={13},
  number={1},
  pages={e91454},
  year={2026},
  publisher={JMIR Publications Inc., Toronto, Canada}
}

@article{ibrahim2026sycophantic,
  title={Sycophantic AI makes human interaction feel more effortful and less satisfying over time},
  author={Ibrahim, Lujain and Hafner, Franziska Sofia and Cheng, Myra and Lee, Cinoo and Anselmetti, Rebecca and Willer, Robb and Rocher, Luc and Yang, Diyi},
  journal={arXiv preprint arXiv:2605.07912},
  year={2026}
}

@article{wei2026cascades,
  title={Cascades of Drift: Mental Health Risks of Prolonged AI Conversations},
  author={Wei, Marlynn H},
  journal={Available at SSRN 6433263},
  year={2026}
}

@article{code2017ethical,
  title={Ethical principles of Psychologist and code of conduct},
  author={Code, AE and Psychologists, Part LXIII},
  journal={Published online},
  year={2017}
}

@book{glaser2017discovery,
  title={Discovery of grounded theory: Strategies for qualitative research},
  author={Glaser, Barney and Strauss, Anselm},
  year={2017},
  publisher={Routledge}
}

@article{saxler2024therapeutic,
  title={Therapeutic alliance in individual adult psychotherapy: a systematic review of conceptualizations and measures for face-to-face-and online-psychotherapy},
  author={Saxler, Eva and Schindler, Theresa and Philipsen, Alexandra and Schulze, Marcel and Lux, Silke},
  journal={Frontiers in psychology},
  volume={15},
  pages={1293851},
  year={2024},
  publisher={Frontiers Media SA}
}

@article{moeseneder2019impact,
  title={Impact of confrontations by therapists on impairment and utilization of the therapeutic alliance},
  author={Moeseneder, Laura and Ribeiro, Eug{\'e}nia and Muran, John Christopher and Caspar, Franz},
  journal={Psychotherapy research},
  volume={29},
  number={3},
  pages={293--305},
  year={2019},
  publisher={Taylor \& Francis}
}

@inproceedings{birhane2024ai,
  title={AI auditing: The broken bus on the road to AI accountability},
  author={Birhane, Abeba and Steed, Ryan and Ojewale, Victor and Vecchione, Briana and Raji, Inioluwa Deborah},
  booktitle={2024 IEEE Conference on Secure and Trustworthy Machine Learning (SaTML)},
  pages={612--643},
  year={2024},
  organization={IEEE}
}

@article{mehta2026dynamics,
  title={The Dynamics of Delusion: Modeling Bidirectional False Belief Amplification in Human-Chatbot Dialogue},
  author={Mehta, Ashish and Moore, Jared and Anthis, Jacy Reese and Agnew, William and Lin, Eric and Yin, Peggy and Ong, Desmond C and Haber, Nick and Dweck, Carol},
  journal={arXiv preprint arXiv:2604.25096},
  year={2026}
}

@article{rathje2025sycophantic,
  title={Sycophantic AI increases attitude extremity and overconfidence},
  author={Rathje, Steve and Ye, Meryl and Globig, Laura and Pillai, Raunak and de Mello, Victoria and Van Bavel, Jay},
  year={2025},
  publisher={OSF}
}

@inproceedings{ojewale2025towards,
  title={Towards AI accountability infrastructure: Gaps and opportunities in AI audit tooling},
  author={Ojewale, Victor and Steed, Ryan and Vecchione, Briana and Birhane, Abeba and Raji, Inioluwa Deborah},
  booktitle={Proceedings of the 2025 CHI Conference on Human Factors in Computing Systems},
  pages={1--29},
  year={2025}
}

@article{weilnhammer2026vulnerability,
  title={Vulnerability-amplifying interaction loops: a systematic failure mode in AI chatbot mental-health interactions},
  author={Weilnhammer, Veith and Hou, Kevin YC and Luettgau, Lennart and Summerfield, Christopher and Dolan, Raymond and Nour, Matthew M},
  journal={arXiv preprint arXiv:2602.01347},
  year={2026}
}

@misc{oduro2026chatbot,
  author       = {Oduro, Serena and Vecchione, Briana and Ye, Meryl and Garofalo, Livia},
  title        = {Protecting the Public from Chatbot Harms: Aligning State Policy with Research},
  year         = {2026},
  publisher    = {Data \& Society Research Institute},
  url          = {https://datasociety.net/points/protecting-the-public-from-chatbot-harms-aligning-state-policy-with-research/}
}

@article{rezaeikhonakdar2023ai,
  title={AI chatbots and challenges of HIPAA compliance for AI developers and vendors},
  author={Rezaeikhonakdar, Delaram},
  journal={Journal of Law, Medicine \& Ethics},
  volume={51},
  number={4},
  pages={988--995},
  year={2023},
  publisher={Cambridge University Press}
}

@article{marks2023ai,
  title={AI chatbots, health privacy, and challenges to HIPAA compliance},
  author={Marks, Mason and Haupt, Claudia E},
  journal={Jama},
  volume={330},
  number={4},
  year={2023}
}

@inproceedings{pi2025interactive,
  title={Interactive AI and Human Behavior: Challenges and Pathways for AI Governance},
  author={Pi, Yulu and Turkay, Cagatay and Bogiatzis-Gibbons, Daniel},
  booktitle={Proceedings of the AAAI/ACM Conference on AI, Ethics, and Society},
  volume={8},
  number={3},
  pages={2016--2029},
  year={2025}
}

@article{mireshghallah2024trust,
  title={Trust no bot: Discovering personal disclosures in human-llm conversations in the wild},
  author={Mireshghallah, Niloofar and Antoniak, Maria and More, Yash and Choi, Yejin and Farnadi, Golnoosh},
  journal={arXiv preprint arXiv:2407.11438},
  year={2024}
}

@inproceedings{king2025user,
  title={User privacy and large language models: An analysis of frontier developers’ privacy policies},
  author={King, Jennifer and Klyman, Kevin and Capstick, Emily and Saade, Tiffany and Hsieh, Victoria},
  booktitle={Proceedings of the AAAI/ACM Conference on AI, Ethics, and Society},
  volume={8},
  number={2},
  pages={1465--1477},
  year={2025}
}

@article{de2025emotional,
  title={Emotional manipulation by AI companions},
  author={De Freitas, Julian and Oguz-Uguralp, Zeliha and Kaan-Uguralp, Ahmet},
  journal={arXiv preprint arXiv:2508.19258},
  year={2025}
}

@inproceedings{jain2026interaction,
  title={Interaction context often increases sycophancy in LLMs},
  author={Jain, Shomik and Park, Charlotte and Viana, Matt and Wilson, Ashia and Calacci, Dana},
  booktitle={Proceedings of the 2026 CHI Conference on Human Factors in Computing Systems},
  pages={1--26},
  year={2026}
}

@book{newhill2003negotiating,
  title={Negotiating the therapeutic alliance: A relational treatment guide},
  author={Newhill, Christina E and Safran, Jeremy D and Muran, J Christopher},
  year={2003},
  publisher={Guilford Press}
}

@book{linehan1993cognitive,
  title={Cognitive-behavioral treatment of borderline personality disorder},
  author={Linehan, Marsha},
  year={1993},
  publisher={Guilford press}
}

@article{susser2019online,
  title={Online manipulation: Hidden influences in a digital world},
  author={Susser, Daniel and Roessler, Beate and Nissenbaum, Helen},
  journal={Geo. L. Tech. Rev.},
  volume={4},
  pages={1},
  year={2019},
  publisher={HeinOnline}
}

@article{fogg2002persuasive,
  title={Persuasive technology: using computers to change what we think and do},
  author={Fogg, Brian J},
  journal={Ubiquity},
  volume={2002},
  number={December},
  pages={2},
  year={2002},
  publisher={ACM New York, NY, USA}
}

@article{zhang2026fragility,
  title={The Fragility of AI Companionship: Ontological, Structural, and Normative Uncertainty in Human-AI Relationships},
  author={Zhang, Renwen and Xie, Lezi},
  journal={arXiv preprint arXiv:2605.03367},
  year={2026}
}

@article{weizenbaum1966eliza,
  title={ELIZA—a computer program for the study of natural language communication between man and machine},
  author={Weizenbaum, Joseph},
  journal={Communications of the ACM},
  volume={9},
  number={1},
  pages={36--45},
  year={1966},
  publisher={ACM New York, NY, USA}
}

@article{li2026into,
  title={Into the black box: Laypeople's folk theories about generative artificial intelligence chatbots},
  author={Li, Zhuoman and Kim, Nuri and Lou, Chen},
  journal={Big Data \& Society},
  volume={13},
  number={2},
  pages={20539517261447838},
  year={2026},
  publisher={SAGE Publications Sage UK: London, England}
}

@article{hatherley2025moving,
  title={A moving target in AI-assisted decision-making: dataset shift, model updating, and the problem of update opacity: J. Hatherley},
  author={Hatherley, Joshua},
  journal={Ethics and Information Technology},
  volume={27},
  number={2},
  pages={20},
  year={2025},
  publisher={Springer}
}

@article{mclaughlin2014patterns,
  title={Patterns of therapeutic alliance: Rupture--repair episodes in prolonged exposure for posttraumatic stress disorder.},
  author={McLaughlin, AnnaMaria Aguirre and Keller, Stephanie M and Feeny, Norah C and Youngstrom, Eric A and Zoellner, Lori A},
  journal={Journal of consulting and clinical psychology},
  volume={82},
  number={1},
  pages={112},
  year={2014},
  publisher={American Psychological Association}
}

@inproceedings{ibrahim2025towards,
  title={Towards interactive evaluations for interaction harms in human-AI systems},
  author={Ibrahim, Lujain and Huang, Saffron and Ahmad, Lama and Bhatt, Umang and Anderljung, Markus},
  booktitle={Proceedings of the AAAI/ACM Conference on AI, Ethics, and Society},
  volume={8},
  number={2},
  pages={1302--1310},
  year={2025}
}

@article{tang2026beyond,
  title={Beyond the Single Turn: Reframing Refusals as Dynamic Experiences Embedded in the Context of Mental Health Support Interactions with LLMs},
  author={Tang, Ningjing and Qian, Alice and Wang, Qiaosi and Howe, Esther and Bullwinkel, Blake and Pedrelli, Paola and Suh, Jina and Heidari, Hoda and Shen, Hong},
  journal={arXiv preprint arXiv:2602.01694},
  year={2026}
}

@article{poonsiriwong2026death,
  title={" Death" of a Chatbot: Investigating and Designing Toward Psychologically Safe Endings for Human-AI Relationships},
  author={Poonsiriwong, Rachel and Archiwaranguprok, Chayapatr and Pataranutaporn, Pat},
  journal={arXiv preprint arXiv:2602.07193},
  year={2026}
}

@misc{openai2025sycophancy,
  title={Sycophancy in GPT-4o: What happened and what we’re doing about it},
  author={OpenAI},
  year={2025},
  publisher={OpenAI: Product}
}

@article{bucher2025s,
  title={“It’s Not Only Attention We Need”: Systematic Review of Large Language Models in Mental Health Care},
  author={Bucher, Andreas and Egger, Sarah and Vashkite, Inna and Wu, Wenyuan and Schwabe, Gerhard},
  journal={JMIR mental health},
  volume={12},
  pages={e78410},
  year={2025},
  publisher={JMIR Publications Toronto, Canada}
}

@article{de2024lessons,
  title={Lessons from an app update at Replika AI: identity discontinuity in human-AI relationships},
  author={De Freitas, Julian and Castelo, Noah and U{\u{g}}uralp, Ahmet K and O{\u{g}}uz-U{\u{g}}uralp, Zeliha},
  journal={arXiv preprint arXiv:2412.14190},
  year={2024}
}

@article{lemmens2021therapy,
  title={Therapy processes associated with sudden gains in cognitive therapy for depression: Exploring therapeutic changes in the sessions surrounding the gains},
  author={Lemmens, Lotte HJM and DeRubeis, Robert J and Tang, Tony Z and Schulte-Strathaus, Julia CC and Huibers, Marcus JH},
  journal={Frontiers in Psychiatry},
  volume={12},
  pages={576432},
  year={2021},
  publisher={Frontiers}
}

@article{kazantzis2006can,
  title={Can between-session (homework) activities be considered a common factor in psychotherapy?},
  author={Kazantzis, Nikolaos and Ronan, Kevin R},
  journal={Journal of Psychotherapy Integration},
  volume={16},
  number={2},
  pages={115},
  year={2006},
  publisher={Educational Publishing Foundation}
}

@article{ardito2011therapeutic,
  title={Therapeutic alliance and outcome of psychotherapy: historical excursus, measurements, and prospects for research},
  author={Ardito, Rita B and Rabellino, Daniela},
  journal={Frontiers in psychology},
  volume={2},
  pages={270},
  year={2011},
  publisher={Frontiers Research Foundation}
}

@article{guo2024large,
  title={Large language models for mental health applications: systematic review},
  author={Guo, Zhijun and Lai, Alvina and Thygesen, Johan H and Farrington, Joseph and Keen, Thomas and Li, Kezhi},
  journal={JMIR mental health},
  volume={11},
  number={1},
  pages={e57400},
  year={2024},
  publisher={JMIR Publications Inc., Toronto, Canada}
}

@article{golden2026transdiagnostic,
  title={A transdiagnostic model for how general purpose AI chatbots can perpetuate OCD and anxiety disorders},
  author={Golden, Ashleigh and Aboujaoude, Elias},
  journal={NPJ Digital Medicine},
  volume={9},
  number={1},
  pages={343},
  year={2026},
  publisher={Nature Publishing Group}
}

@article{guingrich2025longitudinal,
  title={A Longitudinal Randomized Control Study of Companion Chatbot Use: Anthropomorphism and Its Mediating Role on Social Impacts},
  author={Guingrich, Rose E and Graziano, Michael SA},
  journal={arXiv preprint arXiv:2509.19515},
  year={2025}
}

@article{fang2025ai,
  title={How ai and human behaviors shape psychosocial effects of extended chatbot use: A longitudinal randomized controlled study},
  author={Fang, Cathy Mengying and Liu, Auren R and Danry, Valdemar and Lee, Eunhae and Chan, Samantha WT and Pataranutaporn, Pat and Maes, Pattie and Phang, Jason and Lampe, Michael and Ahmad, Lama and others},
  journal={arXiv preprint arXiv:2503.17473},
  year={2025}
}

@article{moore2026characterizing,
  title={Characterizing delusional spirals through human-LLM chat logs},
  author={Moore, Jared and Mehta, Ashish and Agnew, William and Anthis, Jacy Reese and Louie, Ryan and Mai, Yifan and Yin, Peggy and Cheng, Myra and Paech, Samuel J and Klyman, Kevin and others},
  journal={arXiv preprint arXiv:2603.16567},
  year={2026}
}

@inproceedings{moore2025expressing,
  title={Expressing stigma and inappropriate responses prevents LLMs from safely replacing mental health providers.},
  author={Moore, Jared and Grabb, Declan and Agnew, William and Klyman, Kevin and Chancellor, Stevie and Ong, Desmond C and Haber, Nick},
  booktitle={Proceedings of the 2025 ACM Conference on Fairness, Accountability, and Transparency},
  pages={599--627},
  year={2025}
}

@article{zhang2026interaction,
  title={Interaction with AI Companions and Psychological Well-being},
  author={Zhang, Yutong and Zhao, Dora and Hancock, Jeffrey T and Kraut, Robert and Yang, Diyi},
  year={2026},
  publisher={Nature Human Behavior}
}

@article{hoffman2024understanding,
  title={Understanding young adults’ attitudes towards using AI chatbots for psychotherapy: The role of self-stigma},
  author={Hoffman, Benjamin David and Oppert, Michelle Leanne and Owen, Mikaela},
  journal={Computers in Human Behavior: Artificial Humans},
  volume={2},
  number={2},
  pages={100086},
  year={2024},
  publisher={Elsevier}
}

@article{simpson2023differentiating,
  title={Differentiating authentic versus pseudo vulnerability in therapeutic practice},
  author={Simpson, Susan G and Navot, Limor},
  journal={Frontiers in Psychiatry},
  volume={14},
  pages={1200981},
  year={2023},
  publisher={Frontiers Media SA}
}

@article{heinz2025randomized,
  title={Randomized trial of a generative AI chatbot for mental health treatment},
  author={Heinz, Michael V and Mackin, Daniel M and Trudeau, Brianna M and Bhattacharya, Sukanya and Wang, Yinzhou and Banta, Haley A and Jewett, Abi D and Salzhauer, Abigail J and Griffin, Tess Z and Jacobson, Nicholas C},
  journal={Nejm Ai},
  volume={2},
  number={4},
  pages={AIoa2400802},
  year={2025},
  publisher={Massachusetts Medical Society}
}

@article{stubbe2018therapeutic,
  title={The therapeutic alliance: The fundamental element of psychotherapy},
  author={Stubbe, Dorothy E},
  journal={Focus},
  volume={16},
  number={4},
  pages={402--403},
  year={2018},
  publisher={American Psychiatric Association Washington, DC}
}

@article{xu2025digital,
  title={The Digital Therapeutic Alliance With Mental Health Chatbots: Diary Study and Thematic Analysis},
  author={Xu, Zian and Lee, Yi-Chieh and Stasiak, Karolina and Warren, Jim and Lottridge, Danielle},
  journal={JMIR Mental Health},
  volume={12},
  pages={e76642},
  year={2025},
  publisher={JMIR Publications Toronto, Canada}
}

@article{darcy2021evidence,
  title={Evidence of human-level bonds established with a digital conversational agent: cross-sectional, retrospective observational study},
  author={Darcy, Alison and Daniels, Jade and Salinger, David and Wicks, Paul and Robinson, Athena},
  journal={JMIR Formative Research},
  volume={5},
  number={5},
  pages={e27868},
  year={2021},
  publisher={JMIR Publications Toronto, Canada}
}

@misc{woebot2024,
  author       = {{Woebot Health}},
  title        = {Woebot Health},
  year         = {2024},
  url          = {https://woebothealth.com/},
  note         = {Accessed 2025}
}

@misc{wysa2024,
  author       = {{Wysa}},
  title        = {Wysa: {AI} Mental Health Support},
  year         = {2024},
  url          = {https://www.wysa.com/},
  note         = {Accessed 2025}
}

@article{fitzpatrick2017delivering,
  title={Delivering cognitive behavior therapy to young adults with symptoms of depression and anxiety using a fully automated conversational agent (Woebot): a randomized controlled trial},
  author={Fitzpatrick, Kathleen Kara and Darcy, Alison and Vierhile, Molly},
  journal={JMIR mental health},
  volume={4},
  number={2},
  pages={e7785},
  year={2017},
  publisher={JMIR Publications Inc., Toronto, Canada}
}

@techreport{nationalcouncil2024ccbhcimpact,
  author      = {{National Council for Mental Wellbeing}},
  title       = {2024 {CCBHC} Impact Report},
  institution = {National Council for Mental Wellbeing},
  year        = {2025},
  url         = {https://www.thenationalcouncil.org/wp-content/uploads/2025/04/24.06.3_2024-CCBHC-Impact-Report_FINAL.pdf.pdf}
}

@techreport{hrsa2025behavioral,
  author      = {{National Center for Health Workforce Analysis}},
  title       = {State of the Behavioral Health Workforce, 2025},
  institution = {Health Resources and Services Administration, U.S. Department of Health and Human Services},
  year        = {2025},
  month       = {December},
  url         = {https://bhw.hrsa.gov/sites/default/files/bureau-health-workforce/data-research/Behavioral-Health-Workforce-Brief-2025.pdf}
}

@techreport{gao2022mentalhealth,
  author      = {{U.S. Government Accountability Office}},
  title       = {Mental Health Care: Access Challenges for Covered Consumers and Relevant Federal Efforts},
  institution = {U.S. Government Accountability Office},
  year        = {2022},
  month       = {March},
  number      = {GAO-22-104597},
  address     = {Washington, DC},
  url         = {https://www.gao.gov/assets/gao-22-104597.pdf}
}

@misc{de2025disclosure,
  title={Disclosure, humanizing, and contextual vulnerability of generative AI chatbots},
  author={De Freitas, Julian and Cohen, I Glenn},
  journal={NEJM AI},
  volume={2},
  number={2},
  pages={AIpc2400464},
  year={2025},
  publisher={Massachusetts Medical Society}
}

@article{sharma2023towards,
  title={Towards understanding sycophancy in language models, 2023},
  author={Sharma, Mrinank and Tong, Meg and Korbak, Tomasz and Duvenaud, David and Askell, Amanda and Bowman, Samuel R and Cheng, Newton and Durmus, Esin and Hatfield-Dodds, Zac and Johnston, Scott R and others},
  journal={URL https://arxiv. org/abs/2310.13548},
  year={2023}
}

@article{ibrahim2026training,
  title={Training language models to be warm can reduce accuracy and increase sycophancy},
  author={Ibrahim, Lujain and Hafner, Franziska Sofia and Rocher, Luc},
  journal={Nature},
  volume={652},
  number={8112},
  pages={1159--1165},
  year={2026},
  publisher={Nature Publishing Group UK London}
}

@article{cheng2026sycophantic,
  title={Sycophantic AI decreases prosocial intentions and promotes dependence},
  author={Cheng, Myra and Lee, Cinoo and Khadpe, Pranav and Yu, Sunny and Han, Dyllan and Jurafsky, Dan},
  journal={Science},
  volume={391},
  number={6792},
  pages={eaec8352},
  year={2026},
  publisher={American Association for the Advancement of Science}
}

\end{document}